\documentclass[preprint, superscriptaddress, showpacs,preprintnumbers,amsmath,amssymb,prd]{revtex4}
\usepackage{graphicx}
\usepackage{amsmath}\usepackage{slashed}

\begin{document}

\thispagestyle{empty}

\title{Origin of large thermal effect in the Casimir interaction between
two graphene sheets}

\author{
G.~L.~Klimchitskaya}
\affiliation{Central Astronomical Observatory at Pulkovo of the Russian Academy of Sciences, St.Petersburg,
196140, Russia}
\affiliation{Institute of Physics, Nanotechnology and
Telecommunications, St.Petersburg State
Polytechnical University, St.Petersburg, 195251, Russia}

\author{
V.~M.~Mostepanenko}
\affiliation{Central Astronomical Observatory at Pulkovo of the Russian Academy of Sciences, St.Petersburg,
196140, Russia}
\affiliation{Institute of Physics, Nanotechnology and
Telecommunications, St.Petersburg State
Polytechnical University, St.Petersburg, 195251, Russia}

\begin{abstract}
Using the recently derived representation for the polarization tensor
in (2+1)-dimensional space-time allowing an analytic continuation to
the entire plane of complex frequencies, we obtain simple analytic
expressions for the reflection coefficients
of graphene at nonzero Matsubara frequencies. In the framework of
the Lifshitz theory, these coefficients are shown to lead to
nearly exact results for the Casimir free energy and pressure
between two graphene sheets. The constituent parts of large thermal
effect, arising in the Casimir interaction at short separations due
to an explicit parametric dependence of the polarization tensor on
the temperature and an implicit dependence through
a summation over the Matsubara
frequencies, are calculated. It is demonstrated that an explicit
thermal effect exceeds an implicit one at shorter separations,
both effects are similar in magnitudes at moderate separations,
and that an implicit effect becomes a larger one with further
increase of separation. Possible applications of the developed
formalism, other than the Casimir effect, are discussed.
\end{abstract}

\maketitle

\section{Introduction}

Graphene has already assumed great importance as a material possessing
unusual mechanical, electrical and optical properties of high promise
for various applications \cite{1}. These properties are due to 2D
character of graphene. In contrast to standard 3D materials,
quasiparticles in pristine (undoped) graphene are massless Dirac fermions
which possess the linear dispersion relation but move with the Fermi
velocity rather than with the velocity of light \cite{1,2}.
Taking into account that graphene is considered as a potential element
of micro- and nanoelectromechanical devices, much attention is attracted
to forces acting between two graphene sheets and between a graphene sheet
and some element made of an
ordinary material (metallic, dielectric or semiconductor).
These forces are caused by the zero-point and thermal fluctuations of
the electromagnetic field and are called the {\it van der Waals} or
{\it Casimir forces} depending on the range of separations considered
\cite{3,4}. The fluctuation induced forces are known also under the
generic name of {\it dispersion forces}.

The van der Waals and Casimir forces acting between two graphene sheets
and between a graphene sheet and a 3D material were studied theoretically
by many authors in the framework of various approaches (see, for instance,
Refs.~\cite{5,6,7,8,9,10,11,12,12a,13,14}). The most important breakthrough
was made in Ref.~\cite{8}, where it was shown that for graphene the thermal
effect in the Casimir force becomes crucial at by an order of magnitude
shorter separations than for ordinary materials. This result was obtained
using the longitudinal density-density correlation function in the
random-phase approximation, and the relativistic effects were shown to
be not essential.

The most straightforward description of the van der Waals and Casimir
forces in layered systems including graphene is given by the Lifshitz
theory \cite{3,4,15} with reflection coefficients expressed via the
polarization tensor of graphene in (2+1)-dimensional space-time.
The explicit analytic expressions for the components of this tensor
were obtained in Refs.~\cite{16,17} at zero and nonzero temperature,
respectively. Using the values of the polarization tensor of graphene
at the imaginary Matsubara frequencies, a lot of numerical computations
of the Casimir and Casimir-Polder forces between graphene and different
materials and atoms has been performed \cite{16,17,18,19,20,21,22,23,24,25,26}.
The formalisms exploiting the polarization tensor and the density-density
correlation function have been compared \cite{27}. It was shown \cite{27}
that these formalisms are equivalent if one uses the exact
temperature-dependent longitudinal and transverse density-density
correlation functions. Computations using the polarization tensor confirmed
the existence of large thermal effect in the Casimir interaction of
graphene sheets at short separations, as predicted
 in Ref.~\cite{8}. Furthermore, the results computed
using the polarization tensor of graphene at nonzero temperature \cite{28,29}
were found to be in very good agreement with first measurement of the gradient
of the Casimir force between a Au-coated sphere and a graphene-coated
substrate by means of a dynamic atomic force microscope \cite{30}.
The computational results obtained using the hydrodynamic model of graphene
\cite{6,7,11,25} were shown to be excluded by the measurement data over
a wide range of separations \cite{31}.

In this paper, we investigate an origin of the large thermal effect in the
interaction of two graphene sheets at short separations from a few nanometers
to a few hundred nanometers. Our aim is to determine the relative
contributions to the Casimir free energy and pressure due to
 an explicit dependence of the polarization tensor on the
temperature as a parameter and an implicit dependence on the temperature
through  a summation over the Matsubara frequencies.
 For this purpose, we circumvent
cumbersome computations using the exact expressions for the polarization
tensor and find a transparent way of performing simple analytic calculations
in powers of a natural small parameter as far as possible.
It is significant that the polarization tensor of graphene at nonzero
temperature obtained in Ref.~\cite{17} is not well adapted for analytic
calculations. The point is that it is valid only at the discrete imaginary
Matsubara frequencies and its immediate continuation to the entire imaginary
frequency axis does not satisfy necessary physical conditions \cite{32}.
Another form of the polarization tensor of graphene, valid over the whole
plane of complex frequency, including the imaginary and real frequency axes,
was found in Ref.~\cite{32}. At the imaginary Matsubara frequencies the
polarization tensor of Ref.~\cite{32} takes the same values as the one
of Ref.~\cite{17}, but presents better opportunities for analytic
calculations due to much simpler mathematical structure.

Here, we use the polarization tensor in the representation of Ref.~\cite{32}
and obtain simple asymptotic expressions for both the thermal correction
to this tensor and for the reflection coefficients of the electromagnetic
oscillations from a graphene sheet at the Matsubara frequencies.
The developed formalism is applied to calculate the
Casimir free energy and pressure between two graphene sheets at
room temperature.
We compare our results with previously made \cite{21}
numerical computations using exact expressions for the
polarization tensor in the representation of Ref.~\cite{17}.
We demonstrate that
the error of our asymptotic results is only a small fraction of
a percent over the whole region of physically meaningful values of
parameters.  Then we investigate the origin of large thermal
effect at short separations. It is shown that the explicit and implicit
contributions to the dependencies of the Casimir free energy and pressure
 of graphene on the temperature
are equally important parts of the large thermal effect.
In so doing, the major contribution to the thermal effect due to an
explicit temperature dependence of the polarization tensor originates
from the zero-frequency Matsubara term.

The structure of the paper is the following. In Sec.~II we introduce
the main notations, formulate our perturbation approach and obtain
simple asymptotic expressions for the polarization tensor of graphene
and reflection coefficients. Section~III is devoted to the
investigation of large thermal effect in the Casimir free energy of
two parallel graphene sheets. In Sec.~IV the same is done for the
Casimir pressure. Section~V contains our conclusions and discussion.

\section{Simple asymptotic expressions for the polarization tensor of
graphene and reflection coefficients}

In the framework of the Dirac model, the reflection coefficients of the
electromagnetic oscillations from graphene can be expressed via the
polarization tensor of graphene in (2+1)-dimensional space-time.
To calculate the Casimir free energy and pressure between two graphene
sheets one needs the reflection coefficients calculated at the pure
imaginary Matsubara frequencies \cite{17,19,21}
\begin{eqnarray}
&&
r_{\rm TM}(i\xi_l,k_{\bot})=
\frac{q_l\Pi_{00}(i\xi_l,k_{\bot})}{q_l\Pi_{00}(i\xi_l,k_{\bot})+
2\hbar k_{\bot}^2},
\nonumber \\
&&
r_{\rm TE}(i\xi_l,k_{\bot})=-
\frac{\Pi(i\xi_l,k_{\bot})}{\Pi(i\xi_l,k_{\bot})+
2\hbar k_{\bot}^2q_l}
\label{eq1}
\end{eqnarray}
\noindent
for two independent polarizations of the electromagnetic field,
transverse magnetic (TM) and transverse electric (TE).
Here, $\xi_l=2\pi k_BTl/\hbar$, where $k_B$ is the Boltzmann constant,
$T$ is the temperature and $l=0,\,1,\,2,\,\ldots\,$,
are the Matsubara frequencies, $k_{\bot}$ is the magnitude of the
projection of the photon wave vector on the plane of graphene,
$q_l=(k_{\bot}^2+\xi_l^2/c^2)^{1/2}$, $\Pi_{mn}$ with
$m,\,n=0,\,1,\,2$ are the components of the polarization tensor of
graphene in (2+1)-dimensional space-time, and the quantity $\Pi$ is
defined as
\begin{equation}
\Pi(i\xi_l,k_{\bot})=k_{\bot}^2\Pi_{\rm tr}(i\xi_l,k_{\bot})-
q_l^2\Pi_{00}(i\xi_l,k_{\bot}),
\label{eq2}
\end{equation}
\noindent
where $\Pi_{\rm tr}\equiv\Pi_{m}^{\,\,m}$.
Note that the reflection coefficients (\ref{eq1}) can be equivalently rewritten
via the longitudinal and transverse density-density correlation functions, or
conductivities, or polarizabilities or nonlocal dielectric permittivities of
graphene \cite{27}.

The components of the polarization tensor of graphene at temperature $T$
can be presented in the form
\begin{eqnarray}
&&
\Pi_{00}(i\xi_l,k_{\bot})=\Pi_{00}^{(0)}(i\xi_l,k_{\bot})+
\Delta_T\Pi_{00}(i\xi_l,k_{\bot}),
\nonumber \\
&&
\Pi(i\xi_l,k_{\bot})=\Pi^{(0)}(i\xi_l,k_{\bot})+
\Delta_T\Pi(i\xi_l,k_{\bot}),
\label{eq3}
\end{eqnarray}
\noindent
where $\Pi_{00}^{(0)},\ \Pi^{(0)}$ are the respective quantities
calculated at zero temperature, but with continuous frequencies $\xi$ replaced
by the Matsubara frequencies $\xi_l$, and $\Delta_T\Pi_{00}$,
$\Delta_T\Pi$ are the thermal corrections to them.

For a pristine (gapless) graphene under consideration in this paper,
the components of the polarization tensor at $T=0\,$K take especially
simple form \cite{16}
\begin{equation}
\Pi_{00}^{(0)}(i\xi_l,k_{\bot})=\frac{\pi\alpha\hbar k_{\bot}^2}{\tilde{q}_l},
\qquad
\Pi^{(0)}(i\xi_l,k_{\bot})=\pi\alpha\hbar k_{\bot}^2\tilde{q}_l,
\label{eq4}
\end{equation}
\noindent
where $\alpha=e^2/(\hbar c)\approx 1/137$ in the fine-structure constant,
\begin{equation}
\tilde{q}_l=\left(\frac{v_F^2}{c^2}k_{\bot}^2+
\frac{\xi_l^2}{c^2}\right)^{1/2}
\label{eq5}
\end{equation}
\noindent
and $v_F\approx c/300$ is the Fermi velocity.
The respective reflection coefficients (\ref{eq1}) are given by
\begin{equation}
r_{\rm TM}^{(0)}(i\xi_l,k_{\bot})=
\frac{\pi\alpha q_l}{\pi\alpha q_l+2\tilde{q}_l},
\qquad
r_{\rm TE}^{(0)}(i\xi_l,k_{\bot})=-
\frac{\pi\alpha \tilde{q}_l}{\pi\alpha\tilde{q}_l+2{q}_l}.
\label{eq5a}
\end{equation}

For the thermal corrections to the polarization tensor we use the
expressions obtained in Ref.~\cite{32}
\begin{eqnarray}
&&
\Delta_T\Pi_{00}(i\xi_l,k_{\bot})=\frac{16\alpha\hbar c^2}{v_F^2}
\int_{0}^{\infty}\frac{dq_{\bot}}{e^{\hbar cq_{\bot}/(k_BT)}+1}
\nonumber \\
&&~~~~~
\times\left[1-\frac{1}{\tilde{q}_l}{\rm Re} W(\xi_l,k_{\bot},q_{\bot})\right],
\label{eq6} \\
&&
\Delta_T\Pi(i\xi_l,k_{\bot})=\frac{16\alpha\hbar c^2}{v_F^2}
\int_{0}^{\infty}\frac{dq_{\bot}}{e^{\hbar cq_{\bot}/(k_BT)}+1}
\nonumber \\
&&~~~
\times\left\{-\frac{\xi_l^2}{c^2}+{\tilde{q}_l}
\left[
{\rm Re}\, W(\xi_l,k_{\bot},q_{\bot})-\frac{v_F^2k_{\bot}^2}{c^2}
{\rm Re} \frac{1}{W(\xi_l,k_{\bot},q_{\bot})}\right]\right\}.
\nonumber
\end{eqnarray}
\noindent
Here, the integration is performed over the magnitude of a two-dimensional
wave vector of the electronic excitation and the quantity $W$ is defined as
\begin{equation}
 W(\xi_l,k_{\bot},q_{\bot})=
 \left(\tilde{q}_l^2-4q_{\bot}^2+\frac{4i\xi_lq_{\bot}}{c}
 \right)^{1/2}.
 \label{eq7}
 \end{equation}
\noindent
The expressions in Eq.~(\ref{eq6})  have the same values as the respective more
complicated expressions in Ref.~\cite{17}, but, in contrast to the latter,
can be immediately analytically continued to the whole plane of complex
frequency.

It is possible to simplify Eq.~(\ref{eq6}) taking into account that
\begin{eqnarray}
&&
{\rm Re} W(\xi_l,k_{\bot},q_{\bot})=\frac{1}{\sqrt{2}}\left[
\sqrt{(\tilde{q}_l^2-4q_{\bot}^2)^2+\frac{16\xi_l^2q_{\bot}^2}{c^2}}+
\tilde{q}_l^2-4q_{\bot}^2\right]^{1/2},
\nonumber\\
&&
{\rm Re} \frac{1}{W(\xi_l,k_{\bot},q_{\bot})}=
\frac{{\rm Re} W(\xi_l,k_{\bot},q_{\bot})}{\sqrt{(\tilde{q}_l^2-4q_{\bot}^2)^2+
\frac{16\xi_l^2q_{\bot}^2}{c^2}}}.
\label{eq8}
\end{eqnarray}
\noindent
Using Eqs.~(\ref{eq5}) and (\ref{eq8}) and introducing the new integration variable
$u=2q_{\bot}/\tilde{q}_l$, we rewrite Eq.~(\ref{eq6}) in the form
\begin{eqnarray}
&&
\Delta_T\Pi_{00}(i\xi_l,k_{\bot})=\frac{8\alpha\hbar c^2\tilde{q}_l}{v_F^2}
\int_{0}^{\infty}\frac{du}{e^{B_lu}+1}
\nonumber \\
&&~~~~~
\times\left\{1-\frac{1}{\sqrt{2}}\left[
\sqrt{(1+u^2)^2-4\frac{v_F^2k_{\bot}^2u^2}{c^2\tilde{q}_l^2}}+
1-u^2\right]^{1/2}\right\},
\label{eq9} \\
&&
\Delta_T\Pi(i\xi_l,k_{\bot})=\frac{8\alpha\hbar c^2\tilde{q}_l}{v_F^2}
\int_{0}^{\infty}\frac{du}{e^{B_lu}+1}
\nonumber \\
&&~~~~~
\times\left\{
\vphantom{\frac{v_F^2k_{\bot}^2}{c^2\tilde{q}_l^2
\sqrt{(1+u^2)^2-4\frac{v_F^2k_{\bot}^2u^2}{c^2\tilde{q}_l^2}}}}
-\frac{\xi_l^2}{c^2}+\frac{\tilde{q}_l^2}{\sqrt{2}}\left[
\sqrt{(1+u^2)^2-4\frac{v_F^2k_{\bot}^2u^2}{c^2\tilde{q}_l^2}}+
1-u^2\right]^{1/2}\right.
\nonumber \\
&&~~~~~
\times\left.\left[1-\frac{v_F^2k_{\bot}^2}{c^2\tilde{q}_l^2
\sqrt{(1+u^2)^2-4\frac{v_F^2k_{\bot}^2u^2}{c^2\tilde{q}_l^2}}}
\right]
\right\},
\nonumber
\end{eqnarray}
\noindent
where $B_l\equiv\hbar c\tilde{q}_l/(2k_BT)$.

Now we consider approximate expressions for the quantities (\ref{eq9})
applicable at any $l\geq 1$ for calculations of the Casimir force
using the Lifshitz formula.
For this purpose we note that at room temperature
$T=300\,$K the first Matsubara frequency
$\xi_1\approx 2.4\times 10^{14}\,$rad/s. Taking into account that the
characteristic photon wave vector giving the major contribution into the
Casimir free energy and pressure is equal to $k_{\bot}=1/(2a)$, where $a$
is the separation distance between the plates \cite{4,33}, one concludes
that the natural parameter entering Eq.~(\ref{eq9}) can be considered as
a small one at all separations $a\geq 10\,$nm,
\begin{equation}
\frac{4v_F^2k_{\bot}^2}{c^2\tilde{q}_l^2}<
\frac{4v_F^2k_{\bot}^2}{c^2\tilde{q}_1^2}\ll 1.
\label{eq10}
\end{equation}
\noindent
Thus, at $a=10\,$nm this parameter is less than 0.17, at $a=20\,$nm
less than 0.04 and further decreases with increase of separation.

Now we expand the square roots in Eq.~(\ref{eq9}) in powers of the small
parameter (\ref{eq10}) and preserve only the zero and first order contributions.
Furthermore, from Eq.~(\ref{eq10}) we have $v_F^2k_{\bot}^2\ll\xi_1^2$ and,
thus, $B_l\approx\hbar\xi_l/(2k_BT)=\pi l$. Then from Eq.~(\ref{eq9}) one obtains
\begin{eqnarray}
&&
\Delta_T\Pi_{00}(i\xi_l,k_{\bot})=\frac{\alpha\hbar k_{\bot}^2}{\tilde{q}_l}\,Y_l,
\nonumber \\
&& \label{eq11} \\[-6mm]
&&
\Delta_T\Pi(i\xi_l,k_{\bot})={\alpha\hbar k_{\bot}^2}{\tilde{q}_l}Y_l,
\nonumber
\end{eqnarray}
\noindent
where we have introduced the notation
\begin{equation}
Y_l=4\int_{0}^{\infty}\frac{du}{e^{\pi lu}+1}\,\frac{u^2}{1+u^2}.
\label{eq12}
\end{equation}

By combining Eqs.~(\ref{eq4}) and (\ref{eq11}) in accordance with Eq.~(\ref{eq3}),
we arrive to the following simple asymptotic expressions for the polarization
tensor of graphene
\begin{eqnarray}
&&
\Pi_{00}(i\xi_l,k_{\bot})=\frac{\alpha\hbar k_{\bot}^2}{\tilde{q}_l}
(\pi +Y_l),
\nonumber \\
&&
\label{eq13} \\[-6mm]
&&
\Pi(i\xi_l,k_{\bot})=\alpha\hbar k_{\bot}^2\tilde{q}_l
(\pi +Y_l).
\nonumber
\end{eqnarray}
\noindent
Substituting Eq.~(\ref{eq13}) in Eq.~(\ref{eq1}), one obtains simple
equations for the reflection coefficients
\begin{eqnarray}
&&
r_{\rm TM}(i\xi_l,k_{\bot})=
\frac{\alpha q_l(\pi +Y_l)}{\alpha q_l(\pi +Y_l)+2\tilde{q}_l},
\nonumber \\[1mm]
&&
r_{\rm TE}(i\xi_l,k_{\bot})=-
\frac{\alpha \tilde{q}_l(\pi +Y_l)}{\alpha\tilde{q}_l(\pi +Y_l)+2{q}_l}.
\label{eq14}
\end{eqnarray}
\noindent
As can be seen in Eqs.~(\ref{eq13}) and (\ref{eq14}), in both the polarization
tensor and reflection coefficients the relative contribution of each term with
$l\geq 1$ is characterized by the ratio $Y_l/\pi$. In this connection it is
instructive that for $l=1$, 2, 3, 4, and 5 the quantity $Y_l/\pi$ is equal
to 0.041, 0.0074, 0.0024, 0.0011, and $5.5\times 10^{-4}$, respectively,
and further decreases with increasing $l$.

Recall that Eqs.~(\ref{eq13}) and (\ref{eq14}) are valid under a condition $l\geq 1$.
For $l=0$ the thermal correction to $\Pi_{00}$ can be obtained from the first
formula in Eq.~(\ref{eq9})
\begin{equation}
\Delta_T\Pi_{00}(0,k_{\bot})=\frac{16\alpha c}{v_F^2}\left(
k_BT\ln 2-\frac{\hbar v_Fk_{\bot}}{2}
\int_{0}^{1}\frac{du}{e^{B_0u}+1} \sqrt{1-u^2}\right),
\label{eq15}
\end{equation}
\noindent
where $B_0=\hbar v_Fk_{\bot}/(2k_BT)$. In the derivation of Eq.~(\ref{eq15})
we have taken into account that
\begin{equation}
\sqrt{(1-u^2)^2}=\left\{
\begin{tabular}{ll}
$1-u^2$,&\quad $u\leq 1$, \\
$u^2-1$,&\quad $u>1$.
\end{tabular}
\right.
\label{eq16}
\end{equation}
\noindent
Using the first formula in Eq.~(\ref{eq4}), from Eqs.~(\ref{eq1}),  (\ref{eq3})
and (\ref{eq15}) we obtain
\begin{equation}
r_{\rm TM}(0,k_{\bot})=\frac{\alpha c\left\{16k_BT\ln 2+\hbar k_{\bot}v_F[\pi-
8X(k_{\bot})]\right\}}{\alpha c\left\{16k_BT\ln 2+\hbar k_{\bot}v_F[\pi-
8X(k_{\bot})]\right\}+2\hbar k_{\bot}v_F^2},
\label{eq17}
\end{equation}
\noindent
where
\begin{equation}
X(k_{\bot})\equiv\int_{0}^{1}\frac{du}{e^{B_0u}+1} \sqrt{1-u^2}.
\label{eq18}
\end{equation}
\noindent
Equation (\ref{eq17}) is an exact one. It can be rewritten in an equivalent form
\begin{equation}
r_{\rm TM}(0,k_{\bot})=1-\frac{2\hbar k_{\bot}v_F^2}{\alpha c\left\{16k_BT\ln 2+
\hbar k_{\bot}v_F[\pi-8X(k_{\bot})]\right\}+2\hbar k_{\bot}v_F^2}.
\label{eq19}
\end{equation}
\noindent
{}From Eq.~(\ref{eq19}) it is seen that at zero Matsubara frequency the values of the
TM reflection coefficient are rather close to unity and
\begin{equation}
\lim_{k_{\bot}\to 0}r_{\rm TM}(0,k_{\bot})=1.
\label{eq20}
\end{equation}

In a similar way, for $l=0$ the thermal correction to $\Pi$ can be found from
the second formula in Eq.~(\ref{eq9})
\begin{equation}
\Delta_T\Pi(0,k_{\bot})=-\frac{8\alpha \hbar v_Fk_{\bot}^3}{c}\,
Z(k_{\bot}),
\label{eq21}
\end{equation}
\noindent
where
\begin{equation}
Z(k_{\bot})\equiv\int_{0}^{1}\frac{du}{e^{B_0u}+1} \frac{u^2}{\sqrt{1-u^2}}.
\label{eq22}
\end{equation}
\noindent
Using the second formula in Eq.~(\ref{eq4}), from Eqs.~(\ref{eq1}),  (\ref{eq3})
and (\ref{eq21}) one obtains
\begin{equation}
r_{\rm TE}(0,k_{\bot})=-
\frac{\alpha v_F[\pi-8Z(k_{\bot})]}{\alpha v_F[\pi-8Z(k_{\bot})]+2c}.
\label{eq23}
\end{equation}
\noindent
It is easily seen that
\begin{equation}
|r_{\rm TE}(0,k_{\bot}|<\frac{\alpha \pi v_F}{2c}\approx
3.8\times 10^{-5}
\label{eq24}
\end{equation}
\noindent
and, thus, the TE mode does not contribute into the classical limit \cite{22,23}.

\section{Investigation of the thermal effect for the Casimir free energy}

The Casimir free energy per unit area of two parallel graphene sheets separated by
a distance $a$ at temperature $T$ in thermal equilibrium with an environment is
given by the Lifshitz formula \cite{3,4,15}
\begin{eqnarray}
&&
{\cal F}(a,T)=\frac{k_BT}{2\pi}
\sum_{l=0}^{\infty}{\vphantom{\sum}}^{\prime}
\int_{0}^{\infty}k_{\bot}dk_{\bot}\left\{
\ln[1-r_{\rm TM}^2(i\xi_l,k_{\bot})e^{-2aq_l}]\right.
\nonumber \\
&&~~~~~~~~~~~
\left.+
\ln[1-r_{\rm TE}^2(i\xi_l,k_{\bot})e^{-2aq_l}]\right\},
\label{eq25}
\end{eqnarray}
\noindent
where the prime on the summation sign indicates that the term with $l=0$
is divided by two and the reflection coefficients are given in Eq.~(\ref{eq1}).

For convenience in numerical computations we rewrite Eq.~(\ref{eq25}) in terms
of dimensionless variables
\begin{equation}
y=2q_la,\qquad
\zeta_l=\frac{2a\xi_l}{c}.
\label{eq26}
\end{equation}
\noindent
Then the Casimir free energy takes the form
\begin{eqnarray}
&&
{\cal F}(a,T)=\frac{k_BT}{8\pi a^2}
\sum_{l=0}^{\infty}{\vphantom{\sum}}^{\prime}
\int_{\zeta_l}^{\infty}ydy\left\{
\ln[1-r_{\rm TM}^2(i\zeta_l,y)e^{-y}]\right.
\nonumber \\
&&~~~~~~~~~~~
\left.+
\ln[1-r_{\rm TE}^2(i\zeta_l,y)e^{-y}]\right\},
\label{eq27}
\end{eqnarray}
\noindent
where the reflection coefficients are
\begin{eqnarray}
&&
r_{\rm TM}(i\zeta_l,y)=\frac{y\tilde{\Pi}_{00}(i\zeta_l,y)}{y
\tilde{\Pi}_{00}(i\zeta_l,y)+2(y^2-\zeta_l^2)},
\nonumber \\
&&
r_{\rm TE}(i\zeta_l,y)=-\frac{\tilde{\Pi}(i\zeta_l,y)}{\tilde{\Pi}(i\zeta_l,y)
+2y(y^2-\zeta_l^2)}
\label{eq28}
\end{eqnarray}
\noindent
and the dimensionless polarization tensor is given by
\begin{eqnarray}
&&
\tilde{\Pi}_{mn}=\frac{2a}{\hbar}\Pi_{mn},
\label{eq29}\\
&&
\tilde{\Pi}=\frac{8a^3}{\hbar}\Pi=(y^2-\zeta_l^2)\tilde{\Pi}_{\rm tr}-
y^2\tilde{\Pi}_{00}.
\nonumber
\end{eqnarray}

In terms of the variables (\ref{eq26}), the dimensionless polarization tensor at
zero temperature (\ref{eq4}) is equal to
\begin{eqnarray}
&&
\tilde{\Pi}_{00}(i\zeta_l,y)=\frac{\pi\alpha(y^2-\zeta_l^2)}{f_l},
\label{eq30}\\
&&
\tilde{\Pi}(i\zeta_l,y)={\pi\alpha(y^2-\zeta_l^2)f_l},
\nonumber
\end{eqnarray}
\noindent
where
\begin{equation}
f_l\equiv f_l(y)=\left[\tilde{v}_F^2y^2+(1-\tilde{v}_F^2)\zeta_l^2
\right]^{1/2}
\label{eq31}
\end{equation}
\noindent
and $\tilde{v}_F\equiv v_F/c$ is the dimensionless Fermi velocity.
The respective reflection coefficients at $T=0\,$K (where continuous frequency
is replaced with discrete Matsubara frequencies) are obtained from Eq.~(\ref{eq28})
\begin{eqnarray}
&&
r_{\rm TM}^{(0)}(i\zeta_l,y)=\frac{\pi\alpha y}{\pi\alpha y+2f_l},
\nonumber \\
&&
r_{\rm TE}^{(0)}(i\zeta_l,y)=-\frac{\pi\alpha f_l}{\pi\alpha f_l+2y}.
\label{eq31a}
\end{eqnarray}

The dimensionless thermal correction to the polarization tensor (\ref{eq30})
is easily obtained from Eq.~(\ref{eq9})
\begin{eqnarray}
&&
\Delta_T\tilde{\Pi}_{00}(i\zeta_l,y)=\frac{8\alpha f_l}{\tilde{v}_F^2}
\int_{0}^{\infty}\frac{u}{e^{B_lu}+1}
\nonumber \\
&&~~~~~
\times\left\{1-\frac{1}{\sqrt{2}}\left[
\sqrt{(1+u^2)^2-4\frac{\tilde{v}_F^2(y^2-\zeta_l^2)u^2}{f_l^2}}+
1-u^2\right]^{1/2}\right\},
\label{eq32} \\
&&
\Delta_T\tilde{\Pi}(i\zeta_l,y)=\frac{8\alpha f_l}{\tilde{v}_F^2}
\int_{0}^{\infty}\frac{du}{e^{B_lu}+1}
\nonumber \\
&&~~~~~
\times\left\{
\vphantom{\frac{v_F^2k_{\bot}^2}{c^2\tilde{q}_l^2
\sqrt{(1+u^2)^2-4\frac{v_F^2k_{\bot}^2u^2}{c^2\tilde{q}_l^2}}}}
-{\zeta_l^2}+\frac{f_l^2}{\sqrt{2}}\left[
\sqrt{(1+u^2)^2-4\frac{\tilde{v}_F^2(y^2-\zeta_l^2)u^2}{f_l^2}}+
1-u^2\right]^{1/2}\right.
\nonumber \\
&&~~~~~
\times\left.\left[1-\frac{\tilde{v}_F^2(y^2-\zeta_l^2)}{f_l^2
\sqrt{(1+u^2)^2-4\frac{\tilde{v}_F^2(y^2-\zeta_l^2)u^2}{f_l^2}}}
\right]
\right\},
\nonumber
\end{eqnarray}
\noindent
where $B_l=\pi f_l/\tau$ and $\tau\equiv4\pi ak_BT/(\hbar c)$ is
dimensionless temperature.

Using Eqs.~(\ref{eq11}) and (\ref{eq30}), we find approximate expressions for
the polarization tensor in terms of dimensionless variables which are valid
at all nonzero Matsubara frequencies
\begin{eqnarray}
&&
\tilde{\Pi}_{00}(i\zeta_l,y)=\frac{\alpha(y^2-\zeta_l^2)}{f_l}
(\pi +Y_l),
\nonumber \\
&&
\label{eq33} \\[-6mm]
&&
\tilde{\Pi}(i\zeta_l,y)=\alpha(y^2-\zeta_l^2)f_l
(\pi +Y_l),
\nonumber
\end{eqnarray}
\noindent
where $Y_l$ is given by Eq.~(\ref{eq12}).
The respective approximate  reflection coefficients (\ref{eq14}) valid at
$l\geq 1$ take the form
\begin{eqnarray}
&&
r_{\rm TM}(i\zeta_l,y)=
\frac{\alpha y(\pi +Y_l)}{\alpha y(\pi +Y_l)+2f_l},
\nonumber \\[1mm]
&&
r_{\rm TE}(i\zeta_l,y)=-
\frac{\alpha f_l(\pi +Y_l)}{\alpha f_l(\pi +Y_l)+2y}.
\label{eq34}
\end{eqnarray}

Finally, the exact reflection coefficients (\ref{eq19}) and (\ref{eq23}) valid
at zero Matsubara frequency in terms of dimensionless  variables are given by
\begin{eqnarray}
&&
r_{\rm TM}(0,y)=1-\frac{2\pi\tilde{v}_F^2y}{\alpha \left\{8\tau\ln 2+
\pi\tilde{v}_Fy[\pi-8X(y)]\right\}+2\pi\tilde{v}_F^2y},
\nonumber \\
&&
r_{\rm TE}(0,y)=-
\frac{\alpha \tilde{v}_F[\pi-8Z(y)]}{\alpha \tilde{v}_F[\pi-8Z(y)]+2},
\label{eq35}
\end{eqnarray}
\noindent
where the quantities $X(y)$ and $Z(y)$ are defined in Eqs.~(\ref{eq18}) and
(\ref{eq22}) taking into account that $B_0=\pi\tilde{v}_Fy/\tau$.

Now we are in a position to determine the accuracy of our approximate expressions
and to investigate the origin of large thermal effect in the Casimir free energy
between two graphene sheets at short separations. For this purpose we
calculate the Casimir free energy (\ref{eq27}) per unit area of two graphene
sheets in three different ways. First, we compute the exact free energy
${\cal F}(a,T)$ using Eqs.~(\ref{eq27}) and (\ref{eq28}), where the polarization
tensor is given by the sum of respective expressions in Eqs.~(\ref{eq30}) and
(\ref{eq32}). Then, we perform calculations of the approximate free energy
${\cal F}^{(1)}(a,T)$ by using the exact reflection coefficients (\ref{eq35})
at $\zeta_0=0$ and our asymptotic reflection coefficients (\ref{eq34}) at
all $\zeta_l$ with $l\geq 1$. Finally, we calculate the free energy
${\cal F}^{(2)}(a,T)$ using the approach of Ref.~\cite{17}, i.e., by using the
exact reflection coefficients (\ref{eq35}) at $\zeta_0=0$ and
coefficients (\ref{eq31a}) at all $l\geq 1$. Note that the approximate free
energy ${\cal F}^{(1)}(a,T)$ takes an exact account of the explicit temperature
dependence of the polarization tensor in the zero-frequency term of the Lifshitz
formula. In all terms with $l\geq 1$ the explicit temperature dependence is taken
into account approximately using our asymptotic approach. The approximate free
energy ${\cal F}^{(2)}(a,T)$ takes into account an explicit temperature
dependence only in the zero-frequency term and discards it in all terms with
$l\geq 1$.

\begin{figure}[b]
\vspace*{-3cm}
\centerline{\hspace*{1cm}
\includegraphics{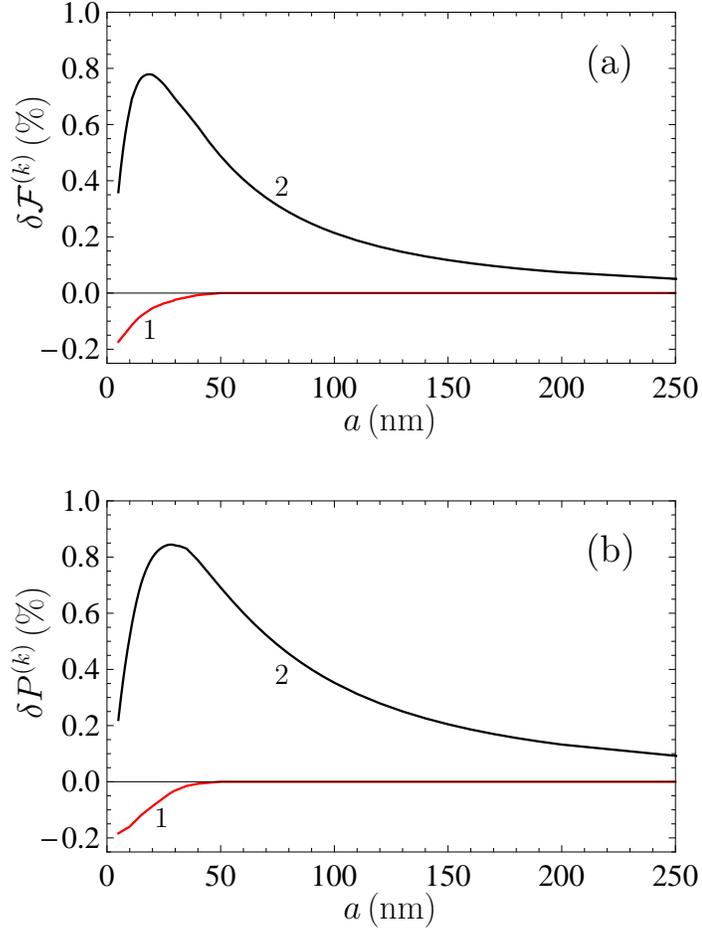}
}
\vspace*{-13cm}
\caption{\label{fg1}(Color online)
The relative errors of the approximate methods for calculation of
the Casimir (a) free energy and (b) pressure
between two graphene sheets using (line 1)
our asymptotic approach taking into account an explicit temperature
dependence of the polarization tensor in all Matsubara terms and
(line 2) discarding this dependence in terms with $l\geq 1$ are
shown as functions of separation.
}
\end{figure}

To compare the accuracies of both approximate calculation approaches, in Fig.~\ref{fg1}(a)
we plot the quantity
\begin{equation}
\delta{\cal F}^{(k)}(a,T)=\frac{{\cal F}(a,T)-{\cal F}^{(k)}(a,T)}{{\cal F}(a,T)}
\label{eq36}
\end{equation}
\noindent
at $T=300\,$K as a function of separation by the lines 1 and 2 in the region from
$a=5$ to 250\,nm ($k=1,\,2$) for the first and second approaches, respectively.
As is seen in Fig.~\ref{fg1}(a) (line 1), the error of our asymptotic approach is
negative and the maximum of its magnitude equal to 0.17\% is achieved at the
shortest separation $a=5\,$nm. With increase of separation to 10 and 30\,nm
the error magnitude decreases to 0.12\% and 0.02\%, respectively.
At separations $a\geq 35\,$nm our calculation approach becomes practically exact.

Note that the characteristic value of $y$ giving the major contribution to the
Casimir free energy (\ref{eq27}) is $y=1$. Taking this into account, the inequality
(\ref{eq10}) reduces to $\tau\gg\tilde{v}_F$. This inequality determines the
region of separations and temperatures where our asymptotic approach is applicable
for calculations of the Casimir free energy and pressure (for the latter see Sec.~IV)
in the framework of the Lifshitz theory. According to the results of Sec.~II,
at $T=300\,$K it should work well at $a\geq 10\,$nm, where the parameter (\ref{eq10})
is less than 0.17. From Fig.~\ref{fg1}(a) (line 1) it is seen that the proposed
asymptotic approach leads to rather accurate result even at $a=5\,$nm where
the parameter (\ref{eq10})  is equal to 0.6. One should take into account,
however, that the Dirac model of graphene is only applicable at energies below
a few eV. This questions the possibility of using it for theoretical description
of the Casimir force at separations below approximately 20 or even 30\,nm \cite{31}.

The second approximate way of calculations turned out to be less accurate than the
first one. As is seen from line 2 in Fig.~\ref{fg1}(a), in this case the calculation
error is positive. It is equal to 0.36\% at $a=5\,$nm, achieves the largest value
of 0.78\% at $a=20\,$nm and then decreases to 0.49\%, 0.21\%, and 0.05\% at
separations 50, 100, and 250\,nm, respectively. Taking into account, however,
that in the second way of calculations the error remains to be below 1\%,
we conclude that the major contribution of an explicit temperature dependence of
the polarization tensor to the Casimir effect originates from the zero-frequency
Matsubara term.

\begin{figure}[b]
\vspace*{-3cm}
\centerline{\hspace*{1cm}
\includegraphics{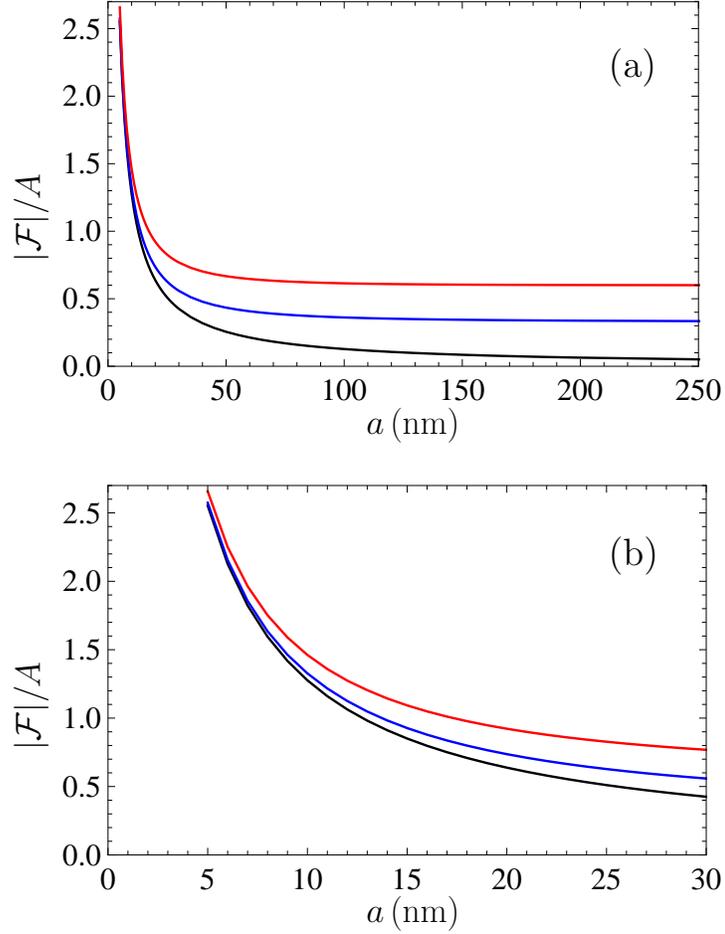}
}
\vspace*{-13cm}
\caption{\label{fg2}(Color online)
The normalized Casimir free energy per unit area of two graphene sheets
at $T=300\,$K calculated (the upper line) exactly and (the intermediate
line) taking into account only an implicit  temperature
dependence through a summation over the Matsubara
frequencies are shown as functions of separation.
The normalized Casimir energy at zero temperature is presented as a
function of separation by the lowest line.
}
\end{figure}

Now we discuss an origin and sources of the large thermal effect in the Casimir
free energy arising for graphene at short separations \cite{8,17,19,21}.
For this purpose in Fig.~\ref{fg2}(a) we plot the magnitudes of the Casimir
free energy normalized to the quantity
$A=k_BT/(8\pi a^2)$ at $T=300\,$K [see Eq.~(\ref{eq27})] as functions of
separation in the region from 5 to 250\,nm. The upper line (${\cal F}$)
was computed at $T=300\,$K using the Lifshitz formula (\ref{eq27}) and the
reflection coefficients (\ref{eq28})  with the exact polarization tensor
defined as a sum of respective expressions in Eqs.~(\ref{eq30}) and (\ref{eq32}).
The nearly coinciding line [${\cal F}={\cal F}^{(1)}$] is obtained also by
using our asymptotic reflection coefficients (\ref{eq34}) at $l\geq 1$ and
(\ref{eq35}) at $l=0$. The intermediate line [${\cal F}^{(0)}$]
was computed at $T=300\,$K using the Lifshitz formula (\ref{eq27}) and the
reflection coefficients (\ref{eq31a}), i.e., using the polarization tensor
at $T=0$, where continuous frequency was replaced by the discrete Matsubara
frequencies. In this case an explicit dependence of the polarization tensor
on $T$ was discarded. Finally, the lowest line (${\cal F}_{0}$) presents the
Casimir free energy at $T=0$ (i.e., the Casimir energy $E(a)$ per unit area
of two graphene sheets) which, by definition, does not contain any thermal
effect, either explicit or implicit.

As can be seen in Fig.~\ref{fg2}(a), there is large implicit thermal effect
for two graphene sheets at short separations due to a summation over
the discrete frequencies. The magnitude of this effect is characterized by
the difference between the intermediate and lowest lines [for better
visualization in Fig.~\ref{fg2}(b) the region from 5 to 30\,nm is shown
separately on an enlarged scale]. The difference between the upper and
intermediate lines characterizes the magnitude of the explicit thermal
effect arising from the dependence of the polarization tensor on the
temperature as a parameter [see also Fig.~\ref{fg2}(b) at shorter
separations]. As to the difference between the upper and lowest lines,
it represents the total thermal correction to the Casimir energy.
As is seen    in Fig.~\ref{fg2}, at moderate separations the explicit and
implicit contributions to the thermal effect are nearly equal.

In Table~I we illustrate the quantitative relationships between different
constituent parts of the thermal correction and the role of the total
thermal correction in the Casimir free energy at $T=300\,$K.
In columns 2, 3, and 4 the fractions of explicit,
$({|{\cal F}|-|{\cal F}^{(0)}|})/{|{\cal F}|}$, implicit,
$({|{\cal F}^{(0)}|-|{\cal F}_0|})/{|{\cal F}|}$, and total
$({|{\cal F}|-|{\cal F}_0|})/{|{\cal F}|}$, thermal corrections are
presented at different separations indicated in column 1.
The quantities $|{\cal F}|$, $|{\cal F}^{(0)}|$ and
${|{\cal F}_0|}=|E|$ are those plotted by the upper, intermediate and
lowest lines in Fig.~\ref{fg2}, respectively.
As is seen in Table~I (column 4), the thermal effect in graphene quickly
increases with increase of separation and provides more than 80\%
contribution to the magnitude of the Casimir free energy at separations
above 100\,nm. In so doing at separations below 200\,nm the TM mode
contributes more than 99\% of the free energy and more than 99.9\% at
$a>200\,$nm. An explicit thermal effect (column 2) contributes more than
one half of the thermal correction at separations below 155\,nm and
at larger separations becomes less than an implicit thermal effect.
However, our calculations show, that the magnitudes of both effects are
similar and, thus, both of them must be taken into account in
computations. We note also the role of the Matsubara term with $l=0$.
At separation distances of 30, 100, 200, and 1000\,nm it contributes
69.3\%, 94.4\%, 98.4\%, and 99.94\% of the Casimir free energy,
respectively.

As was mentioned in Sec.~I, the thermal effect in the Casimir force
for graphene becomes dominant at much shorter separations than for
plates made of metallic or dielectric materials. For example,
for ideal metal plates the relative thermal correction to the
Casimir free energy at $T=300\,$K is equal to
$(|{\cal F}|-|{\cal F}_0|)/|{\cal F}|=3\times 10^{-5}$ and
$2.6\times 10^{-2}$ at separation distances $a=100$ and 1000\,nm,
respectively. This should be compared with Table~I, where the
respective relative thermal effects for graphene are indicated as
0.7922 and 0.9788, i.e., by the factors of $2.6\times 10^5$ and
37.6 larger.

The physical reason for such a big difference between graphene and
other materials is the following. For ordinary metals and dielectrics
the thermal effect becomes dominant when the contribution of all
Matsubara terms with $l\geq 1$ becomes exponentially small due
to the exponential factors in Eq.~(\ref{eq25}).
At $T=300\,$K this happens at separations above 5 or $6\,\mu$m.
In doing so, the remaining zero-frequency Matsubara term alone
describes the classical thermal Casimir effect.
As to graphene, at short separations above 100\,nm, where the
exponential factors in Eq.~(\ref{eq25}) are not small yet,
the reflection coefficients in all Matsubara terms with $l\geq 1$
become very small making dominant the classical thermal
contribution of the zero-frequency term.

\section{Investigation of the thermal effect for the Casimir pressure}

The Lifshitz formula for the Casimir pressure between two parallel graphene
sheets is given by \cite{3,4,15}
\begin{eqnarray}
&&
{P}(a,T)=-\frac{k_BT}{\pi}
\sum_{l=0}^{\infty}{\vphantom{\sum}}^{\prime}
\int_{0}^{\infty}q_lk_{\bot}dk_{\bot}\left\{
[r_{\rm TM}^{-2}(i\xi_l,k_{\bot})e^{2aq_l}-1]^{-1}\right.
\nonumber \\
&&~~~~~~~~~~~
\left.+
[r_{\rm TE}^{-2}(i\xi_l,k_{\bot})e^{2aq_l}-1]^{-1}\right\},
\label{eq38}
\end{eqnarray}
\noindent
where the exact reflection coefficients are written in Eq.~(\ref{eq1})
with the polarization tensor presented in Eqs.~(\ref{eq3}), (\ref{eq4})
and (\ref{eq6}).

In terms of the dimensionless variables (\ref{eq26}) convenient in
numerical computations Eq.~(\ref{eq38}) takes the form
\begin{eqnarray}
&&
{P}(a,T)=-\frac{k_BT}{8\pi a^3}
\sum_{l=0}^{\infty}{\vphantom{\sum}}^{\prime}
\int_{\zeta_l}^{\infty}y^2dy\left\{
[r_{\rm TM}^{-2}(i\zeta_l,y)e^{y}-1]^{-1}\right.
\nonumber \\
&&~~~~~~~~~~~
\left.+
[r_{\rm TE}^{-2}(i\zeta_l,y)e^{y}-1]^{-1}\right\},
\label{eq39}
\end{eqnarray}
\noindent
where the exact reflection coefficients can be found in Eq.~(\ref{eq28})
and the dimensionless  polarization tensor is presented in Eqs.~(\ref{eq30})
and (\ref{eq32}). The approximate reflection coefficients at all nonzero
Matsubara frequencies ($l\geq 1$) using our asymptotic approach are
given in Eq.~(\ref{eq34}). At zero Matsubara frequency
the exact reflection coefficients are written in Eq.~(\ref{eq35}).

First, we determine the accuracy of our approximate method in  application
to calculations of the Casimir pressure. To do this, we
calculate the Casimir pressure (\ref{eq39}) in three different ways
already discussed in Sec.~III. As  a result, we obtain
 the exact Casimir pressure
${P}(a,T)$,  the approximate Casimir pressure
${P}^{(1)}(a,T)$ computed using our asymptotic approach and
the approximate pressure ${P}^{(2)}(a,T)$ using the approach of Ref.~\cite{17}
 discarding an explicit dependence of the polarization tensor on $T$ in
 Matsubara terms with $l\geq 1$.
The accuracies of both approximate methods are characterized by
the quantities
\begin{equation}
\delta{P}^{(k)}(a,T)=\frac{{P}(a,T)-{P}^{(k)}(a,T)}{{P}(a,T)}.
\label{eq40}
\end{equation}

In Fig.~\ref{fg1}(b) we plot the quantities $\delta P^{(1)}$ and $\delta P^{(2)}$
at $T=300\,$K as a function of separation by the lines 1 and 2, respectively.
As is seen in Fig.~\ref{fg1}(b) (line 1), the error of our asymptotic approach
in calculations of the Casimir pressure is
negative and the maximum of its magnitude is equal to 0.18\%  at
$a=5\,$nm. At separations of 10 and 30\,nm
the error magnitude decreases to 0.16\% and 0.01\%, respectively, and becomes nearly zero
at $a\geq 35\,$nm.

For the second approximate way of calculations [the line 2 in Fig.~\ref{fg1}(b)] the
error is positive. It is equal to 0.22\% at $a=5\,$nm, achieves the largest value
of 0.87\% at $a=30\,$nm and then decreases to 0.69\%, 0.35\%, and 0.09\% at
separations 50, 100, and 250\,nm, respectively.
Thus, our asymptotic expressions for the polarization tensor of graphene and
reflection coefficients lead to more exact results than the approximate method
discarding an explicit temperature dependence of the polarization tensor at all
$l\geq 1$. Nevertheless, the main contribution to the explicit thermal effect
in the Casimir pressure, similar to the free energy, originates from the
zero-frequency Matsubara term.

\begin{figure}[b]
\vspace*{-3cm}
\centerline{\hspace*{1cm}
\includegraphics{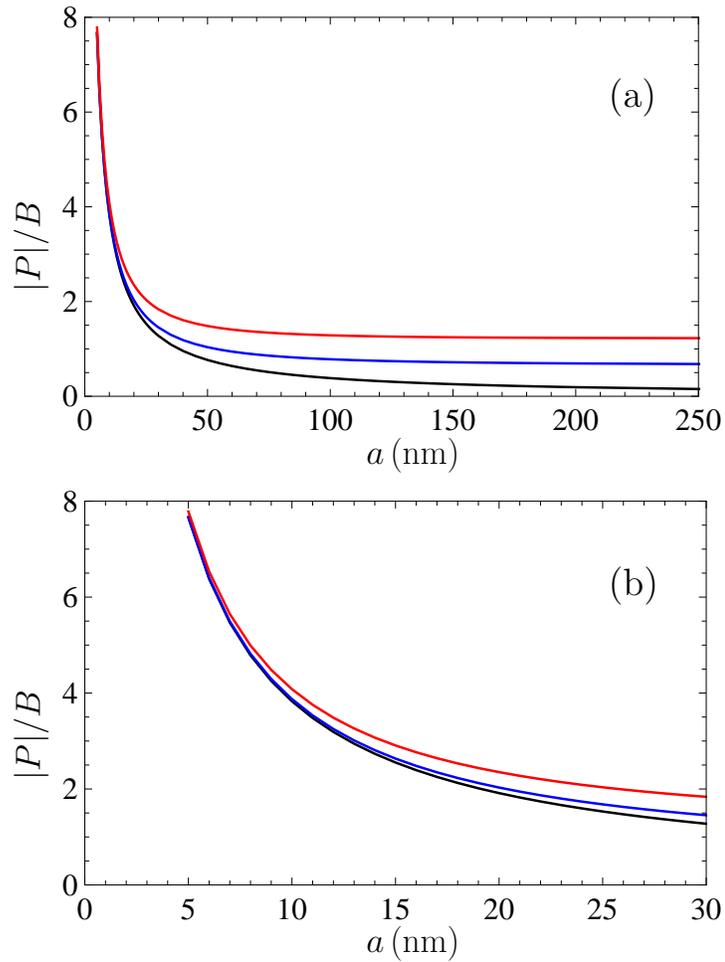}
}
\vspace*{-13cm}
\caption{\label{fg4}(Color online)
The normalized Casimir pressure between two graphene sheets
at $T=300\,$K calculated (the upper line) exactly and (the intermediate
line) taking into account only an implicit  temperature
dependence through a summation over the Matsubara
frequencies are shown as functions of separation.
The normalized Casimir pressure at zero temperature is presented as a
function of separation by the lowest line.
}
\end{figure}

To investigate an origin and sources of the large thermal effect,
in Fig.~\ref{fg4}(a) we plot the magnitudes of the Casimir
pressure normalized to the quantity
$B=k_BT/(8\pi a^3)$ at $T=300\,$K [see Eq.~(\ref{eq27})] as functions of
separation. The upper, intermediate and lowest lines in Fig.~\ref{fg4}(a,b)
were obtained in the same ways as the respective lines in Fig.~\ref{fg2}(a,b),
i.e., the upper line [${P}=P^{(1)}$]
was computed at $T=300\,$K exactly by using Eqs.~(\ref{eq28}), (\ref{eq30}),
(\ref{eq32}) and (\ref{eq39}), the intermediate line [$P^{(0)}$]
at $T=300\,$K  by  Eqs.~(\ref{eq31a}) and (\ref{eq39}), and
the lowest line (${P}_{0}$) is the
Casimir pressure computed  at $T=0$.

{}From Fig.~\ref{fg4}(a) it is seen that, similar to the free energy, there is large
implicit thermal effect in the Casimir pressure between
two graphene sheets at short separations,  characterized by
the difference between the intermediate and lowest lines.
[The separation region from 5 to 30\,nm is shown
 on an enlarged scale in Fig.~\ref{fg4}(b).] The explicit thermal
effect in the Casimir pressure is shown in Fig.~\ref{fg4}(a,b)
by the difference between the upper and
intermediate lines. This contribution to the total thermal correction
(the latter is shown by  the difference between the upper and lowest lines)
originates from the dependence of the polarization tensor on the
temperature as a parameter. Again, at moderate separations the explicit and
implicit thermal effects contribute to the total thermal correction nearly equal.

The quantitative results are listed in Table~II. The first column of this table
contains the values of separation and the columns 2, 3, and 4 the values of
$({|{P}|-|{P}^{(0)}|})/{|{P}|}$,
$({|{P}^{(0)}|-|{P}_0|})/{|{P}|}$, and
$({|{P}|-|{P}_0|})/{|{P}|}$,
where the magnitudes of the Casimir pressure
$|{P}|$, $|{P}^{(0)}|$, and ${|{P}_0|}$ are the exact ones
(the upper line), the ones computed  at $T=300\,$K using the zero-temperature
polarization tensor (the intermediate line), and the ones computed at $T=0$
(the lowest line),  respectively.
Similar to the case of the free energy,
the thermal effect in the Casimir pressure quickly
increases with increase of separation (column 4). At separations
above 150\,nm  the thermal effect contributes more than 80\%
of the Casimir pressure between two graphene sheets.
The contribution of an explicit thermal effect in the Casimir pressure
(column 2) is larger than that of an implicit at all separations below
or equal 325\,nm.
At separation distances 30, 100, 200, and 1000\,nm the zero-frequency
Matsubara term contributes
55.7\%, 89.9\%, 96.9\%, and 99.86\% of the Casimir pressure,
respectively.
Similar to the case of the Casimir free energy, the  thermal effect
for the Casimir pressure becomes dominant at much shorter separations
than for metallic and dielectric materials. Thus, for two ideal metal
plates at separation distances $a=100$ and 1000\,nm one has
$(|P|-|P_0|)/|P|=1.57\times 10^{-7}$ and $1.57\times 10^{-3}$,
respectively. {}From Table~II we conclude that for graphene the relative
thermal effect at the same respective separations is larger by the
factors of $5.6\times 10^7$ and $6.2\times 10^2$.
As to the relative role of the TM and TE modes
in the total Casimir pressure, the TM mode remains dominant and
provides more than 99\% of the pressure magnitude. In precise
computations, however, both modes should be taken into account.

\section{Conclusions and discussion}

In the foregoing we have obtained simple asymptotic expressions for the
polarization tensor of graphene and reflection coefficients at nonzero
Matsubara frequencies suitable for calculation of the Casimir free energy
and pressure between two parallel graphene sheets. This has been made
possible by applying the new form of the polarization tensor \cite{32} valid
not only at the imaginary Matsubara frequencies, but over the entire plane
of complex frequency. In this paper, the analytic asymptotic expressions
are derived by means of perturbation expressions in powers of a natural
parameter which becomes small over the wide range of
physically interesting separations and temperatures.  The respective
expressions for the Casimir free energy and pressure between
two graphene sheets take into
account both an explicit dependence of the polarization tensor (or,
equivalently, conductivity, or dielectric permittivity, or density-density
correlation function) on the temperature as a parameter and an implicit
temperature dependence through a summation over the Matsubara frequencies.

Using the obtained asymptotic expressions for the reflection coefficients,
we have compared the Casimir free energy and pressure
calculated in the framework of our formalism with the previously known
ones \cite{21} calculated numerically using the polarization tensor
of Ref.~\cite{17}.
It was shown that our asymptotic approach
leads to very accurate results (the maximum errors are equal to 0.17\% and
0.18\% at the separation of 5\,nm for the Casimir free energy and pressure,
respectively). These errors decrease with increase of separation. At all
separations above 35\,nm the asymptotic Casimir free energies and pressures
coincide with the exact ones.

We have investigated an origin of large thermal effect arising in the Casimir
interaction between graphene sheets at short separations \cite{8}.
We have shown that the total thermal corrections to the Casimir energy and
pressure at zero temperature consist of two parts.
The explicit part is determined by the parametric dependence
of the polarization tensor on $T$, whereas the implicit one arises through
a summation over the Matsubara frequencies. According to our results,
in both the Casimir free energy and pressure, an explicit thermal effect
exceeds an implicit one at shorter separations. At moderate separations both
parts of the thermal effect are similar in magnitudes. With further increase
of separation an implicit thermal effect in the Casimir free energy and
pressure becomes larger than an explicit one. We have also confirmed
the previously known result \cite{17,21} that
the thermal effect in the Casimir interaction between two graphene sheets
contributes more than 80\% of the Casimir free energy and pressure at
relatively short separations exceeding 100 and 150\,nm, respectively, and
that the major contribution to both the free energy and pressure is given
by the electromagnetic field with transverse magnetic polarization.

In future it would be interesting to extend our asymptotic approach to
real frequency axis and obtain simple analytic expressions for the
polarization tensor, conductivity, density-density correlation functions and
dielectric permittivity of graphene at nonzero temperature for numerous
applications other than the Casimir effect. As an example, one could
mention calculation of the reflectivity properties of graphene over a wide
range of real frequencies.
%%%%%%%%%%%%%%%%%%%%%%%%%%%%%%%%%%

%%%%%%%%%%%%%%%%%%%%%%%%%%%%%%%%%

%%%%%%%%%%%%%%%%%%%%%%%%%%%%%
%%%%%%___Table_I___%%%%%%%%%%%%%%%%%%%
\begingroup
\squeezetable
\begin{table}
\caption{Different constituent parts of the relative thermal correction
to the Casimir free energy (column 4) arising (column 2) from an
explicit dependence of the polarization tensor on temperature and
(column 3) from an implicit temperature dependence through
a summation over the Matsubara frequencies are shown at $T=300\,$K at
different separations between two graphene sheets (column 1).
See text for further discussion.
}
\begin{ruledtabular}
\begin{tabular}{cccc}
$a\,$(nm)&$~~\vphantom{\int\limits_{0}^{0}}
\frac{|{\cal F}|-|{\cal F}^{(0)}|}{|{\cal F}|}~~$ &
$~~\frac{|{\cal F}^{(0)}|-|{\cal F}_0|}{|{\cal F}|}~~$ &
$~~\frac{|{\cal F}|-|{\cal F}_0|}{|{\cal F}|}~~$\\
\hline
5&  0.0312 & 0.0082 & 0.0394\\
10& 0.0922 & 0.0347 & 0.1269\\
20& 0.2013 & 0.1068 & 0.3081\\
30& 0.2730 & 0.1734 & 0.4464\\
50& 0.3492 & 0.2680 & 0.6172\\
100& 0.4115 & 0.3807 & 0.7922\\
150& 0.4303 & 0.4289 & 0.8592\\
200& 0.4387 & 0.4552 & 0.8939\\
250& 0.4432 & 0.4717 & 0.9149\\
500& 0.4511 & 0.5064 & 0.9575\\
1000& 0.4543 & 0.5245 & 0.9788
%\hline
\end{tabular}
\end{ruledtabular}
\end{table}
\endgroup
%%%%%%%%%%%%%%%%%%%%%%%
%%%%%%___Table_II___%%%%%%%%%%%%%%%%%%%
\begingroup
\squeezetable
\begin{table}
\caption{Different constituent parts of the relative thermal correction
to the Casimir pressure (column 4) arising (column 2) from an
explicit dependence of the polarization tensor on temperature and
(column 3) from an implicit temperature dependence through
a summation over the Matsubara frequencies are shown at $T=300\,$K at
different separations between two graphene sheets (column 1).
See text for further discussion.
}
\begin{ruledtabular}
\begin{tabular}{cccc}
$a\,$(nm)&$~~\vphantom{\int\limits_{0}^{0}}
\frac{|{P}|-|{P}^{(0)}|}{|{P}|}~~$ &
$~~\frac{|{P}^{(0)}|-|{P}_0|}{|{P}|}~~$ &
$~~\frac{|{P}|-|{P}_0|}{|{P}|}~~$\\
\hline
5&  0.0144 & 0.0019 & 0.0163\\
10& 0.0506 & 0.0123 & 0.0619\\
20& 0.1365 & 0.0493 & 0.1858\\
30& 0.2086 & 0.0967 & 0.3053\\
50& 0.3014 & 0.1822 & 0.4836\\
100& 0.3928 & 0.3096 & 0.7024\\
150& 0.4234 & 0.3716 & 0.7950\\
200& 0.4374 & 0.4072 & 0.8446\\
250& 0.4452 & 0.4300 & 0.8752\\
500& 0.4584 & 0.4790 & 0.9374\\
1000& 0.4637 & 0.5051 & 0.9688
%\hline
\end{tabular}
\end{ruledtabular}
\end{table}
\endgroup
%%%%%%%%%%%%%%%%%%%%%%%
%%%%%%%%%%%%%%%%%%%%%%%%%%%%%
%%%%%____FIGURE__1___%%%%%%%%%%%%%%%%%%%%%
%%%%%%%%%%%%%%
%%%%%____FIGURE__2___%%%%%%%%%%%%%%%%%%%%%
%%%%%%%%%%%%%%
%%%%%____FIGURE__3___%%%%%%%%%%%%%%%%%%%%%
%%%%%%%%%%%%%%

%%%%%%%%%%%%%%%%%%%%%%%%%%%%%%
\end{document}